\documentclass[sigconf,nonacm]{acmart}
\usepackage{graphicx}
\usepackage[toc]{appendix}
\usepackage{epstopdf}
\usepackage{graphics}
\usepackage{url}
\usepackage{xcolor}
\usepackage{comment}
\usepackage{subcaption}

\AtBeginDocument{%
\providecommand\BibTeX{{%
\normalfont B\kern-0.5em{\scshape i\kern-0.25em b}\kern-0.8em\TeX}}}

\graphicspath{{figures}}

\begin{document}

\title{Supervised similarity learning for corporate bonds using Random Forest proximities}

\author{Jerinsh Jeyapaulraj}
\email{jerinsh.jeyapaulraj@blackrock.com}
\affiliation{
\institution{BlackRock, Inc.}
\country{ New York, NY, USA.}
}

\author{Dhruv Desai}
\email{dhruv.desai1@blackrock.com}
\affiliation{
\institution{BlackRock, Inc.}
\country{ New York, NY, USA.}
}

\author{Peter Chu}
\email{peter.chu@blackrock.com}
\affiliation{
\institution{BlackRock, Inc.}
\country{ New York, NY, USA.}
}

\author{Dhagash Mehta}
\email{dhagash.mehta@blackrock.com}
\affiliation{
\institution{BlackRock, Inc.}
\country{ New York, NY, USA.}
}

\author{Stefano Pasquali}
\email{stefano.pasquali@blackrock.com}
\affiliation{
\institution{BlackRock, Inc.}
\country{ New York, NY, USA.}
}

\author{Philip Sommer}
\email{philip.sommer@blackrock.com}
\affiliation{
\institution{BlackRock, Inc.}
\country{ New York, NY, USA.}
}
\renewcommand{\shortauthors}{Jeyapaulraj et al.}

\begin{abstract}
Financial literature consists of ample research on similarity and comparison of financial assets and securities such as stocks, bonds, mutual funds, etc. However, going beyond correlations or aggregate statistics has been arduous since financial datasets are noisy, lack useful features, have missing data and often lack ground truth or annotated labels. However, though similarity extrapolated from these traditional models heuristically may work well on an aggregate level, such as risk management when looking at large portfolios, they often fail when used for portfolio construction and trading which require a local and dynamic measure of similarity on top of global measure. In this paper we propose a supervised similarity framework for corporate bonds which allows for inference based on both local and global measures. From a machine learning perspective, this paper emphasis that random forest (RF), which is usually viewed as a supervised learning algorithm, can also be used as a similarity learning (more specifically, a distance metric learning) algorithm. In addition, this framework proposes a novel metric to evaluate similarities, and analyses other metrics which further demonstrate that RF outperforms all other methods experimented with, in this work. 
\end{abstract}

\ccsdesc[500]{Applied computing~Economics}

\keywords{Corporate Bonds, Similarity Learning, Random Forest, Proximity, Distance Metric Learning}

\maketitle
\section{Introduction}
Similarity learning has found many applications across different industries ranging from facial recognition, recommendation systems for movies, songs, house price estimates, friendship recommendations on social networks, etc. \cite{aggarwal2016recommender} For financial companies, similarity learning can be used to identify similarity across/within assets, similar investors based on their demographics and transaction histories, similar funds and hedge funds, similarity amongst companies based on their transcript calls or annual reports, similarity amongst portfolio managers, health-state similarity for retirement planning, etc. \cite{lamby2018classifying,hirano2019related,ito2020learning,marathe1999categorizing,haslem2001morningstar,mehta2020machine,satone2021fund2vec,desai2021robustness,thompson2021know,pagliaro2021investor,tan2022health}. 

In the present work, we focus on corporate bonds similarity. The corporate bond market is highly diverse, with hundreds of thousands of securities traded in varying volumes and frequencies. This gives rise to varying liquidity profile for each and often a portion of them is low on liquidity or ill-liquid (SEC Rule 22e-4). The limited dealer inventory hampers the ability of traders/portfolio managers (PMs) to engage in the market.In return, such trades could accumulate, which could lead to higher tracking errors (specifically for index products) and performance impacts (for both index funds and active funds).

A rigorous similarity framework for corporate bonds can be leveraged by traders in the following way. When a bond is to be traded but isn't liquid enough, the trader or PM can scan through bonds which are similar to the original security and choose one which offers a better price or a specific trading characteristic. A tree-based approach will allow us to incorporate a large number of features and capture changes in market dynamic. The similarity derived from this approach will also have captured measures which can provide relative ranking based on these characteristics. Characteristics such as higher liquidity score, better bid or offer price can help the trader make informed decisions. This framework can also be leveraged to replace existing bonds with products which satisfy better attributes such as higher Environmental, Social and Governance (ESG) score.

Machine learning research to identify similarity or clusters for financial assets has primarily focused on stocks, mutual funds and hedge funds (see, e.g., ~\cite{marathe1999categorizing,haslem2001morningstar,das2003hedge,miceli2004ultrametricity,baghai2005consistency,gibson2007style,shawky2010stylistic,mehta2020machine,desai2021robustness,satone2021fund2vec}). While the research on these topics for corporate bonds has been sparse. Ref.~\cite{bagde2018comprehensive} used various clustering algorithms such as K-means, Hierarchical clustering, Self-Organizing Maps, Fuzzy-C Mean and Gaussian mixture model to identify investment grade bonds and high yield bonds based on coupon rate, yield-to-maturity, current yield, credit rating, price and the bond being callable or not. The clusters are then analyzed using ground truth labels that indicate successful and unsuccessful term deposit contracts when telemarketing phone calls were made to Portuguese retail bank clients.

In Ref.~\cite{cole2022trade}, trade prices of corporate bonds were clustered, and it was shown that regardless of size of the trade, both buy and sell dealer trades with customers (relative to interdealer trading) lead to an increase in price clustering while dealers may use clustered prices when purchasing from and selling to institutions (i.e., may use a clustered price to insulate themselves from the risk of asymmetric information).

In Ref.~\cite{wu2020thesis}, corporate bonds from the used dataset were first manually sorted based on their sectors. This data was then clustered using bond attributes such as industry classification, country, bid spread, current yield, yield-to-maturity, option-adjusted spread, option-adjusted duration, outstanding amount, coupon rate, excess return, maturity date, issue date, closing price, credit rating and various key rate durations. For each sector, first the dimensionality of the data was reduced using either multiple factor analysis, principal component analysis (PCA) or singular value decomposition. Then, K-means and hierarchical clustering were used to cluster the bonds in the reduced dimensional space. Using the elbow curve method optimal clusters for each sector were then identified. Finally, to compute the weight of each variable used in clustering, the author posed the problem as a supervised classification problem where the aforementioned variables were used as input features and the cluster numbers of each data-point was used as the target variable. A random forest model was then trained on this data and the weights for each feature were assigned based on feature importance from random forest model.

\subsection{Shortcomings of Unsupervised Learning Methods for Similarity Computations}
Most of the research on similarity learning in the financial domain has been focused on unsupervised methods such as K-means or any other clustering methods on the original datasets or in the lower-dimensional space obtained using PCA or auto-encoders or else.
Here, for the given dataset without any target variable, one performs clustering of data-points such that groups of data-points that are ``similar", as defined explicitly (by feeding a specific distance metric such as Euclidean, Chebyshev, Minkowski, etc., into the model, e.g., K-means) or implicitly supplied in the form of the chosen objective function. The optimal number of clusters is chosen with the help of another metric such as the elbow curve, Silhoutte score, Gap statistics, etc. Then, for a given data-point one can identify the closest ones according to this distance.

First, in a completely unsupervised formulation of a similarity problem, there is no unique metric to evaluate if the result of similarity is ``better'' than any other method or not. The aforementioned metrics only evaluate the algorithms with a corresponding definition of ``quality of clusters". Moreover, a priori one may not know which features are important or even sufficient to define similarity nor the weights or feature importance of the individual features. In addition, here, the manually supplied distance metric may or may not be appropriate for the underlying data manifold. e.g., if the underlying data manifold is a sphere, then the Euclidean distance may not be an appropriate metric.

\subsection{Supervised Similarity Learning}
Ideally, the set of important features as well as importance for each feature should be learned in an algorithmic way using the available data rather than manually supplied. Such a machine learning method that learns the distance metric from the given data is called similarity learning. The main goal of a similarity learning method is to learn the similarity metric that quantifies similarity between pairs of data-points. More specifically, the similarity metric is usually defined as the inverse of a distance metric, and hence, broadly speaking, similarity learning is also known the distance metric learning (DML).

DML is usually performed in a supervised or semi-supervised fashion \footnote{Though there exist several unsupervised learning methods to learn a underlying low-dimensional manifold where distance (or any other chosen geometric properties) between the given data-points are preserved, e.g., principal component analysis, deep autoencoders, ISOMAPs, etc., strictly speaking, they are manifold learning methods \cite{izenman2012introduction} and not necessarily distance metric learning methods.}: for the given data, one retrieves (either explicitly given, or else) labels for pairs of data-points yielding whether the two data-points are close (or, similar) or not. Then, an optimization problem is formulated to reverse engineer the distance metric that would have made the pair-wise labels possible. In other words, instead of relying on off-the-shelf distance metrics, DML aims to algorithmically construct task-specific distance metric from the given data. Such a distance metric learned from the data can then be used as a definition of similarity for the given data; can improve the performance of a distance based model such as the K-nearest neighbor (KNN); improve the performance of a distance-based clustering method such as K-means clustering; to perform semi-supervised tasks; for dimensionality reduction tasks; etc.~\cite{yang2006distance,kulis2012metric,DBLP:journals/corr/abs-1812-05944}.

Starting from some of the earliest works (e.g., \cite{10.5555/2968618.2968683}), one of the most common basic strategies is to start with a parametric distance metric, most commonly the Mahalanobis metric \cite{mahalanobis1936generalized}, and learning the numerical values of the parameters from the available data by solving an optimization problem. Some examples of Mahalanobis distance metric based DML methods are Mahalanobis Metric for Clustering \cite{10.5555/2968618.2968683}, information theoretic metric learning \cite{davis2007information}, DML with Eigenvalue Optimization \cite{ying2012distance}, etc. which learn global distance metric and may ignore the local geometry of the data manifold. Other DML methods that learn distance metric based on local neighborhoods of data-points include Large margin nearest neighbor method \cite{weinberger2009distance}, neighborhood component analysis \cite{goldberger2004neighbourhood}, probabilistic approaches such as \cite{yang2006efficient} etc. There also exist deep learning based DML methods such as Siamese networks \cite{bromley1993signature,chicco2021siamese}. The reader is referred to Refs.~\cite{DBLP:journals/corr/abs-1812-05944} for a recent review on DML methods and Ref.~\cite{kaya2019deep} for a recent review on deep learning based DML methods.

In the aforementioned research as well as review articles, a powerful DML method usually goes unmentioned or only mentioned in passing\footnote{See, e.g., Refs~\cite{pang2006pathway,resende2018survey,zhao2016propensity}, etc. for other examples of applications of RF based similarity learning. }, although it was already known to Breiman himself \cite{breiman-cutler-RF}, that is, the very well-known Random Forest (RF) algorithm\cite{breiman_random_2001}. As briefly reviewed in the next Section, this method is not only efficient for metric learning from large and potentially imbalanced datasets and missing values, it also learns a  local distance metric, i.e., distance metric that depends on the position of the data-points in the space.

In the present work, we argue that the RF method is a powerful method to learn similarity of financial assets in a supervised fashion. To the best of our knowledge, a DML (though a global one) was explicitly used for the first time in Ref.~\cite{desai2021robustness} to learn similarity among mutual funds. Here, we argue that RF is a powerful method to learn similarity from as complex dataset as of corporate bonds. In the next Section, we briefly review the literature on similarity learning using RF, and describe its salient features compared to other DML methods. Then, we apply the RF to learn local distance metric for corporate bond data. Furthermore, we propose a novel objective metric to evaluate DML based similarity methods and apply it in our case. We then discuss the implications of our proposed framework and conclude.

\section{Random Forest as Similarity Learning Method}
In this Section, we describe how RF can be viewed as a similarity learning method, the precise definition of similarity extracted from RF and the salient features of RF as a similarity learning method.

\subsection{Random Forest}
We first describe Decision Tree (DT) which is also a popular and powerful machine learning algorithm for both classification and regression tasks \cite{hastie2009random}. Here, based on the chosen cost function (usually mean squared error for the regression task and Gini or information entropy for the classification task). The algorithm tries different split points for each of the features to identify the most cost efficient (minimum cost) split. Feature which gives the most cost-efficient split is then the topmost level of the corresponding ``decision tree". Then, for each of the branches of the split at this level, once again goes through each feature and each split to identify the next cost-efficient split. One iterates the same splitting process for the remaining features, known as greedy and recursive binary splitting. The algorithm stops when a stopping criterion is met, e.g., a predetermined number of levels (called the depth of the tree), or when minimum number of data-points falling in each branch is reached.

Finally, we obtain terminal branches corresponding to the rectangle partitions of the feature space. Continuing with the tree analogy, the final branches are called the \textit{leaf} nodes, and the path from the top level to a leaf node is called the \textit{decision path} for the leaf node. Then, for every data-point that falls into the same leaf node, we make the same prediction which is equal to the mean (for the regression task) or majority voting (for the classification task) of the target values for the training dataset in the leaf node.

Now, we move to describing RF \cite{hastie2009random,liaw2002classification, breiman2001random} which is an ensemble learning method based on DTs. In RF, one constructs multiple DTs in the following way: first, from given training data, one creates $T$ number of datasets using bootstrapping of the training data with replacement. Then, for each of the $T$ datasets, one randomly selects a subset of input features and constructs a DT using the aforementioned DT algorithm. Eventually, an aggregation of all these smaller trees is used to obtain a final model, justifying the name ''Random Forest''. Here, again, the depth of each DT, the number of DTs, the number of features in each randomly selected subset, etc., are hyperparameters and can be tuned to improve learning. 

RF evades the problem of overfitting, that an individual DT may likely suffer from, due to ensembling of DTs. Hence, RF is usually more accurate, and the generalization error for RF reduces as the number of DTs increases. The generalization error of RF also depends on the strength of the individual DTs in RF and the correlation between them. At the same time RF's are less interpretable due to the presence of multiple small DTs.

\subsection{Random Forest as an Adaptive Nearest-Neighbors Method}
Though a strong relationship between nearest neighbor algorithms and RF may not be immediately obvious, it was indeed shown in Ref.~\cite{lin2006random} that RF is an adaptive nearest-neighbor algorithm. In this reference, first, a generalization of KNN, called potential nearest neighbors (k-PNNs), was introduced. Then, it was shown that RF can be viewed as an adaptively weighted k-PNN method.

\begin{figure*}[htb]
\begin{subfigure}[h]{0.4\textwidth}
\includegraphics[width=\linewidth]{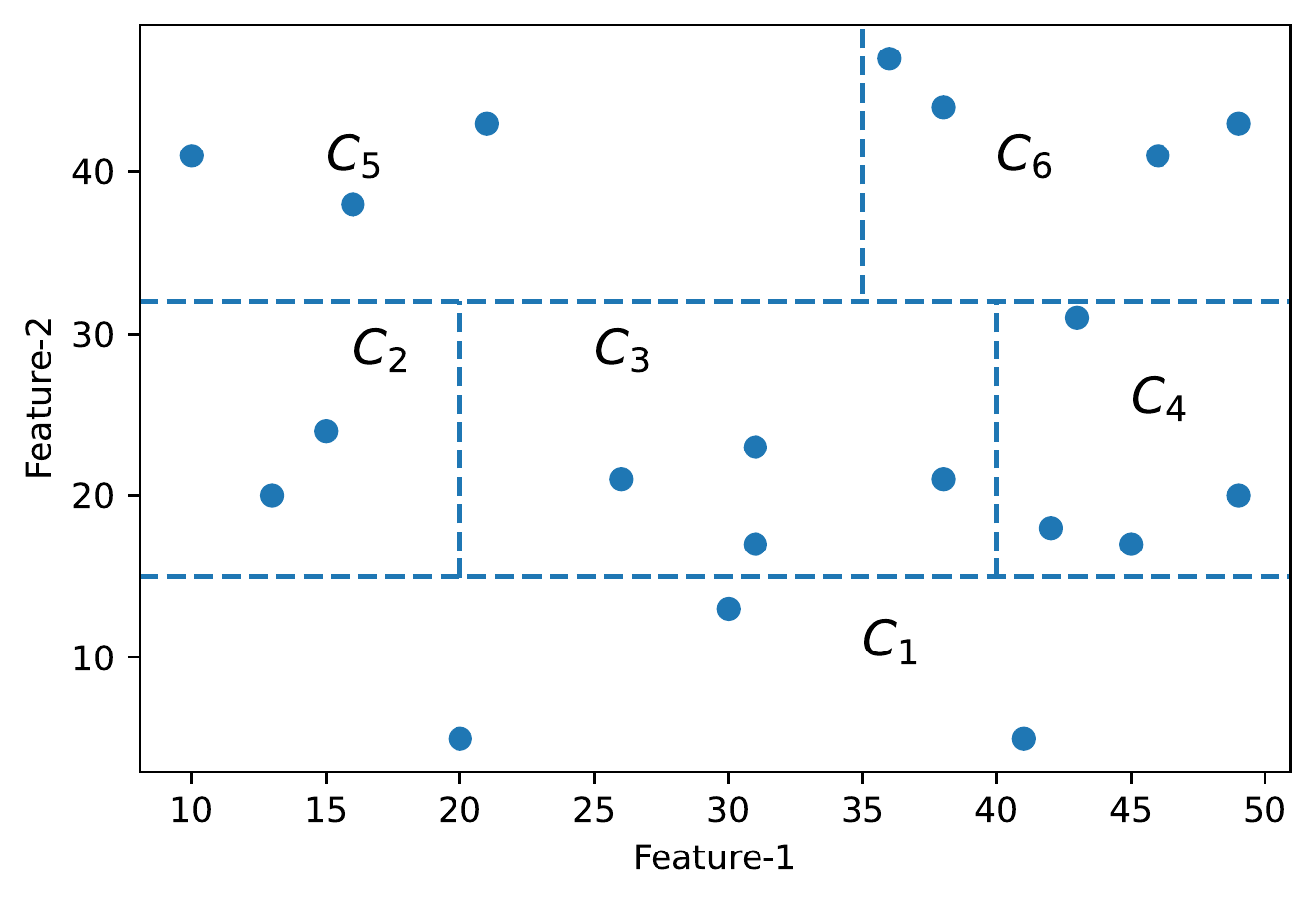}
\caption{Decision Tree partition space}
\label{fig:DT_figure}
\end{subfigure}
~
\begin{subfigure}[h]{0.4\textwidth}
\includegraphics[width=\linewidth]{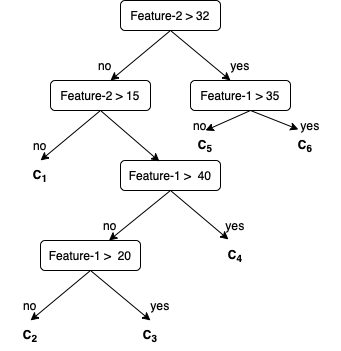}
\caption{Decision Tree splits}
\label{fig:DT_figure_flow}
\end{subfigure}
\caption{Example of decision tree which captures 6 disjoint regions $\mathcal{C}_1, \mathcal{C}_2,...\mathcal{C}_3$ to capture overall feature space from feature-1 and feature-2 as seen in ( \subref{fig:DT_figure}). (\subref{fig:DT_figure_flow}) represents corresponding leaf nodes in decision tree with respect to this disjoint space. For a test point which falls in $\mathcal{C}_1$, its neighbours are all other data-points which belong to $\mathcal{C}_1$. \cite{devavrat2018}}
\end{figure*}

Intuitively, in the regression setting, for example, a leaf node in each DT corresponds to a rectangle partition in the feature space and the predicted value of the target variable for a data-point falling in this partition is the average of the target values of all the training data-points within the partition (see Figure \ref{fig:DT_figure}). In other words, all the training data-points present in the partition can be viewed as the nearest-neighbors of any other data-point within the partition. For a typical DT, different partitions corresponding to different leaf nodes may be of different sizes, meaning that the number of nearest-neighbors is different depending on which partition the data-point lands on. The different size (volume) of the partitions also suggests an underlying non-trivial distance metric learned by the algorithm. Eventually, for RF, one aggregates over all the partitions in all the DTs the given data-point lands on, including additional weights into the aggregation. The same intuition can be extended to the classification setting only by replacing averaging of the target values of the training-data in the partition with majority voting of the target values of the training-data.

In practice, there are multiple ways to compute pair-wise similarity (and, in turn, distance), called \textit{proximity}, using RF proposed in the literature (see, e.g., \cite{englund2012novel,cao2021novel,rhodes2022geometry}). We make use of two traditional ways to compute proximity between each pair of data-points using RF. We will investigate other proximity definitions in the future. Below, we describe the two variants of proximity and provide explicit expressions of the same.

\subsubsection{Original definition of Proximity}
The earliest definition of proximity was provided in \cite{breiman-cutler-RF}: after all the $T$ trees are grown, i.e., a random forest is trained, drop each of the data-points (from both training and OOB sets) down each of the trees. Then, if a pair of data-points fall in the same leaf node, increase their similarity score (i.e., proximity) by one for each tree. Hence, the maximum proximity between a pair of data-points can be $T$ (i.e., the pair of data-points fall in the same leaf node in each of the $T$ trees) and the minimum can be $0$ (i.e., the pair never fall in the same leaf node in any of the $T$ trees). Finally, we normalize the proximities for each pair of data-points by dividing them by $T$. Mathematically, the proximity between data-point $i$ and $j$ is given by 
\begin{equation}
Prox(i,j) = \frac{1}{T} \sum_{t=1}^{T}{I(j \in v_i(t))},
\label{eq:prox}
\end{equation}
where $T$ is the number of trees in the forest, $v_i(t)$ contains indices of the data-points that end up in the same leaf node as $x_i$ in tree $t$, and $I(.)$ is the indicator function. 

\subsubsection{Out-of-Bag Proximity}
The aforementioned definition of proximity uses both in-bag and out-of-bag data-points. By design, the in-bag data-points of different classes will terminate in different leaf nodes if all trees are grown such that each leaf node is pure. This will lead to over-exaggerated class separation in proximity values. One way to resolve this issue is to redefine proximity measure between data-points $i$ and $j$ using only those trees for which both $i$ and $j$ are out-of-bag samples \cite{hastie2009random,liaw2002classification}. Mathematically, this proximity, called out-of-bag (OOB) proximity, is defined as:
\begin{equation}
Prox_{OOB}(i,j) = \frac{\sum_{t\in S_i} I (j \in O(t) \cap v_i(t))}{\sum_{t \in S_i}I(j \in O(t))},
\label{eq:oob_prox}
\end{equation}
where $O(t)$ is the set of indices of OOB training data-points and $S_i$ is the set of trees where observation $i$ is OOB and $v_i$(t) contains indices of all data-points that end up in the same terminal node as data-point i in tree t.

\subsection{Salient Features of Random Forest as a Similarity Method}
Using RF as a DML method brings in all the advantages of RF over other ML methods and beyond:
\begin{itemize}
\item RF is a non-parametric method requiring to tune minimal number of hyperparameters;
\item Unlike many of the traditional DML methods, RF requires minimal amount of data cleaning or data preprocessing. It can even handle missing values at least in theory, handles imbalanced datasets, as well as both numerical and categorical types of input data.
\item RF can handle fairly large datasets making it a scalable DML technique unlike many other traditional methods; 
\item Since RF can handle high-dimensional datasets, there is no specific need to first reduce the dimensionality of the data using principle component analysis or some other methodology unlike many other DML methods; 
\item One of the most important theoretical advantage of using RF as a DML method over traditional methods is that the feature importance of individual features in computing similarity is algorithmically learned by RF for each data-point. A little thought would reveal that the previous statement yields that RF learns local distance metric (i.e., the distance metric that depends on the location in the space) as opposed to only global distance metric.
\end{itemize}

\section{Data Description}
The universe that we tested on was a subset of the global corporate bond market, consisting mostly of US securities (\~10k securities). For this analysis we used the cross-sectional information for the bonds as on 2021-01-01. The cross-sectional features include \cite{sommer2016liquidity}:

\noindent\textbf{Bond Duration:} Measures the sensitivity of the bond's price to changes in the interest rate;

\noindent\textbf{Coupon:} Annual interest paid on the bond expressed as a percentage of the bond's face value;

\noindent\textbf{Coupon Frequency:} Indicates the number of times the coupon is paid in a given year

\noindent\textbf{Country:} The country where the bond is issued;

\noindent\textbf{Age:} The number of days since the bond's issuance;

\noindent\textbf{Days to Maturity:} The number of days to the bond's maturity;

\noindent\textbf{Amount Issued:} The total issue size of the bond offering;

\noindent\textbf{Amount Outstanding:} The total size of the bond that haven't yet matured or been redeemed;

\noindent\textbf{Issuer:} The company that issued the bond to borrow money;

\noindent\textbf{Industry:} The industry of the bond's issuer;

\noindent\textbf{Bond Rating:} A composite rating for the bonds based on the ratings given by various credit rating agencies;

Here, we are interested in identifying bonds which are similar in terms of the above input features with the target variable being related to liquidity. Liquidity of an asset is termed as a \textit{multi-dimensional beast} in Ref.~\cite{sommer2016liquidity}. Leaving technicalities aside, there are four major variables that indicate corporate bond liquidity: bid-ask spreads, yield or return spreads, trading volume, and the price impact by trading (with respect to an appropriate index, for example). In the present work, we use the yield of the bond as given by the latest market price available to us as the target variable to demonstrate a first application of the proposed methodology. The range of the target variable is $[ 0.32,7.82]$. In Table \ref{tab:features_list}, we summarize the variable list and the data type.

\subsection{Data Pre-processing}
We used one-hot-encoding for categorical features. There were no missing values in training and testing data. The transformed dataset consisted of 378 categorical encoded features and 9 numerical features resulting in total of 387 features. 

\begin{table}[]
\begin{tabular}{l l l}
\toprule
Column & Type \\ \hline
\hline
Yield to Maturity (Target Variable) & Numerical \\
Coupon & Numerical \\
Coupon Frequency & Numerical\\
Duration & Numerical \\
Country & Category \\
Days to Maturity & Numerical \\
Age & Numerical \\
Industry & Category \\
Amount Issued & Numerical \\
Amount Outstanding & Numerical\\
Bond Rating & Category \\
\bottomrule
\end{tabular}
\caption{List of target variable and input features with their variable types.}\label{tab:features_list}
\vspace{-6mm}
\end{table}
\raggedbottom

\section{Methodology}\label{sec:methodology}
In this Section, we describe our overall methodology to train and evaluate the RF model on the given data, compute similarities from the final RF model, and eventually evaluate the similarities.

\subsection{Train-test Split and Cross-validation}
We randomly shuffled the dataset and then selected 90\% of the data for training and the remaining 10\% data for testing. Given the size of the data, this was an optimal split size.

For the regression task at hand, we used 5-fold cross validation on the training data to perform hyperparameter optimization and extract the best performing model based on optimal parameters. For creating the 5 folds, we randomly shuffle the training dataset and split it into 5 parts.

\subsection{Metrics to Evaluate Random Forest Results}
For the supervised regression problem at hand, we use mean squared error (MSE) and mean absolute percentage error (MAPE).

\subsubsection{Mean Squared Error}
The MSE of an estimator measures the average of the squared difference between the actual and predicted values, i.e., if $y_i$ and $\hat{y}_i$ are the observed and predicted values of the target variable for the $i$-th data-point, then 
\begin{equation}
\mbox{MSE} = \frac{1}{n} \sum_{i=1}^n (y_i - \hat{y_i})^2.
\end{equation}
Here, smaller value of MSE means a better model.
\subsubsection{Mean Absolute Percentage Error}
The MAPE is the mean or average of the absolute percentage difference between the actual and predicted values \cite{mape}, i.e.,
\begin{equation}
MAPE = \frac{1}{n} \sum_{i=1}^n \left\lvert\frac{y_i-\hat{y_i}}{y_i} \right\rvert.
\end{equation}
Again, smaller MAPE corresponds to a better model.

\subsection{Hyperparameter Optimization}
We performed hyperparameter optimization\cite{probst2019hyperparameters} for the number of trees (estimators) and the max depth until which each tree is grown in the RF. While tuning these hyperparameters we used bootstrap aggregation and kept the following parameters as constant: maximum features used to decide for splits (all features), squared error to calculate quality of split, minimum samples needed for a split being 2. We varied maximum depth from 2 to 25 (with varying step size), and also tested the case of no limit for the depth. For number of estimators we varied them from 100 to 2000 with a step size of 250. In practice, in order to evaluate the performance of the model at these parameter values we used grid search cross-validation as implemented in Scikit-Learn. 

\subsection{Proximity Computations and Similarity Matrix}
Using the trained model, we first extract the OOB and in-bag indices for the training data which were used to train individual tree in RF so that we can actively track OOB samples for each tree. The general idea for proximity and OOB-proximity for training data is to calculate the proximities as defined by \ref{eq:prox} and \ref{eq:oob_prox}, respectively, between data-points $i$ and $j$. 

In case of test data, we calculate the OOB proximity by taking pairs of data-points between train and test data, where the data-point from the training data is OOB. Finally, we generate an $n\times n$ matrix which captures proximity for each data-point against all others. The proximity for a specific data-point with itself will always be $1$, i.e the diagonal of the $n\times n$ matrix.

\subsection{Metrics to Evaluate Similarity}
In general, it is always difficult to evaluate similarity, and especially to determine whether results of one similarity method is \textit{better} than those achieved by another method, as there is no ground truth. In Ref.~\cite{englund2012novel}, the authors had introduced a novel way to compare their proposed definition of RF proximity to the traditional definition: they used the proximity matrices arising from different definitions of RF proximities as kernel matrices in a support vector machine by assuming that a data proximity matrix of higher quality leads to higher classification accuracy. In Ref.~\cite{rhodes2022geometry}, the authors computed the training and testing errors with respect to the observed and proximity-weighted values of the target variables (say, the proximity-weighted training and testing errors). Then, they compared these two errors with the usual training and testing errors with respect to the observed and the RF predicted values of the target variables (say, RF training and testing errors), respectively, for different proximities. Finally, for whichever definition of proximity the corresponding proximity-weighted errors were closer to the RF errors was considered more accurate (i.e., geometry and accuracy preserving) proximity.

In the present paper, we propose another objective evaluation methodology to compare results from different similarity methods. We follow up from the intuition that a better similarity method should bring similar data-points closer to each other. Hence, the performance of the KNN method using the respective distance metric should yield higher performance for the optimal value of K than the distance metric learned from another method.

Here, we compare the performances of KNN with following distance metrics: 
\begin{enumerate}
\item Euclidean distance ($d_E$); 
\item Gower distance ($d_G$) (which is a traditional distance metric for mixed-variable type such as our dataset); 
\item Proximity distance, $d_{Prox}$, where
\begin{equation}
d_{Prox}(i,j) = 1- Prox(i,j);
\end{equation}

\item OOB distance, $d_{Prox_{OOB}}$, where
\begin{equation}
d_{Prox_{OOB}}(i,j) = 1 - Prox_{OOB}(i,j).
\end{equation}
\end{enumerate}

To correctly take the proximity-weighted distances of the data-points into account, when using python's Sklearn \cite{sklearn_api} package for KNN, the implementation allows users to supply a pre-computed distance metric. Here, a pre-computed distance metric can be an explicitly coded distance metric function or an $n\times n$ square matrix which captures the distance from each training point to another. For any test data-point, the user needs to input an $m\times n$ matrix which captures the distance between $m$ test data-points with all the original $n$ training data-points which were used to train the KNN algorithm for that specific distance metric. With this implementation we used distance weighted KNN as compared to uniformly weighted.

In summary, for each of the above distance metrics, we run KNN regression for the same dataset with exactly the same train-test split, for various values of K to identify the optimal value of K. Then, we compare the RMSE and MAPE for the optimal values of K to identify the best distance metric.

\section{Computational Details and Results}
In this Section, we provide computational details, and present the results to summarize our experiments as well as their interpretations. We present results of the training of RF to demonstrate how well the RF was able to reproduce the training and test ground-truth labels. Then, we provide results from the similarity computations using the trained RF.

\subsection{Computational Details}
For this research we used 32-core 32 GB RAM virtual instance on a cloud platform. RF and KNN implementations from sklearn library in python scaled well when used with high number of processors and high memory. Calculating similarity matrix took 10 mins. Storing and calculating proximity matrix requires both efficient computation and large memory because we must take a pass at all leaf nodes of each tree present in the RF for each of the $n \choose 2$ pairs. To the best of our knowledge there are no standard packages in python which provide proximity and OOB proximity calculations from a tree-based method, and hence we created our own which we plan to release in the future.

\subsection{Supervised Learning using Random Forest}
RF was trained on the training split with cross-validation as described in Section \ref{sec:methodology}. To evaluate the model performance we use RMSE and MAPE as our metrics.

\subsubsection{Hyperparameter Optimization and Metrics}
\begin{figure}[h]
\centering
\includegraphics[width=\linewidth]{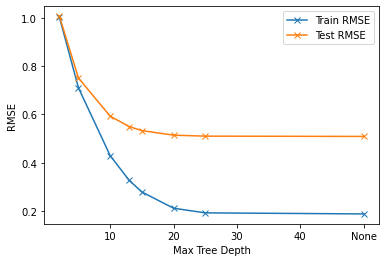}
\caption{Random Forest Hyperparameter (estimators=1000)}
\label{fig:hyperparam}
\end{figure}

Figures \ref{fig:hyperparam} shows RMSE as a function of maximum depth as a hyperparameter, for training and testing splits. Based on these plots, we chose max depth 10 as the optimal parameter to trade-off between bias and variance. 

\begin{table}[htbp]
\begin{tabular}{l l l}
\toprule
Dataset & RMSE & MAPE \\ \hline
\hline
Train & 0.21 & 0.08 \\
Test & 0.51 & 0.15 \\
\bottomrule
\end{tabular}
\caption{RMSE and MAPE for the trained RF model on the train and test datasets.}
\label{tab:rmse_mape}
\vspace{-6mm}
\end{table}

In Table \ref{tab:rmse_mape}, we record the training and testing values of both the metrics. In summary, the trained RF model learned the ground truth function from the data with fairly high accuracy as measured by both the metrics. The high test accuracy also means that the available input features are able to predict the target variable reasonably well, in turn assuring that these input features are indeed sufficient enough for further similarity computations for the specific problem as formulated. Moreover, RF has also captured the (local) importance of each variable for individual predictions which will be implicitly fed into the proximity based similarity computations. Appendix \ref{appendix:a} provides additional details on feature importance for random forest model.

\subsection{Proximity Similarity and Evaluation of Similarity}
\begin{figure}[h]
\centering
\includegraphics[width=\linewidth]{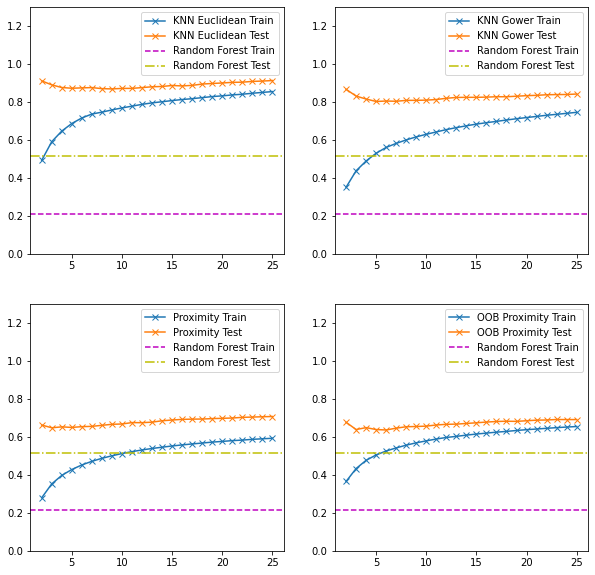}
\caption{KNN RMSE error compared to Random Forest}
\label{fig:RMSE}
\end{figure}

\begin{figure}[h]
\centering
\includegraphics[width=\linewidth]{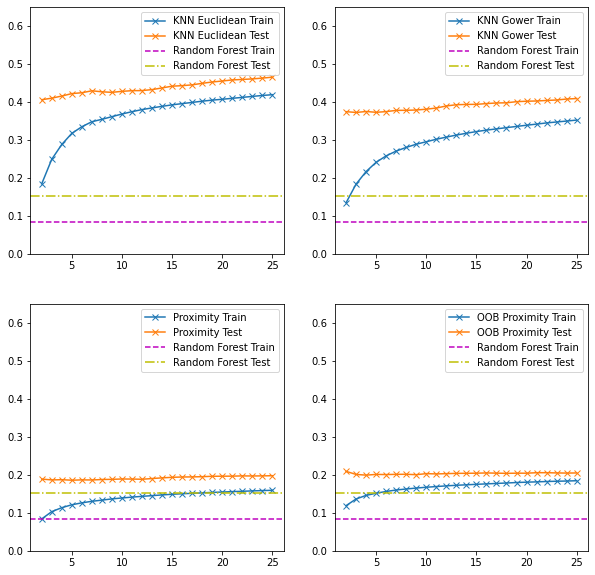}
\caption{KNN MAPE error compared to Random Forest}
\label{fig:MAPE}
\end{figure}

Figures \ref{fig:RMSE} and \ref{fig:MAPE} shows train and test RMSE and MAPE, respectively, for KNN regression with respect to all four distance metrics ($d_E$, $d_G$, $d_{Prox}$ and $d_{Prox_{OOB}}$) for a range of values for K. For comparison, we also plot the train and test RMSEs and MAPEs for the base RF model which appear as constant lines in the figure as they are independent of K. Here, the test RMSE and MAPE for $d_{Prox}$ and $d_{Prox_{OOB}}$ are always lower than $d_E$ and $d_G$ yielding that, on an average, the pairs of data points in the test dataset are correctly identified as 'nearest-neighbors' as per the latter two distance metrics.

We stress that ideally the difference between the RF train (test) RMSE (MAPE) and KNN train (test) RMSE (MAPE) with either $d_{Prox}$ or $d_{Prox_{OOB}}$ should be the same up to statistical fluctuations. The difference seen in these figures suggest that the definitions of proximities do not capture every detail the RF has learned. Recently, Ref.~\cite{rhodes2022geometry} has proposed a so-called Geometry and Accuracy Preserving proximity definition for RF which, in theory, should close the above-mentioned gap. We plan to explore this and other definitions existing in the literature in the future.

\section{Discussion and Conclusion}
Asset similarity has numerous applications in finance in general and in investment processes. For corporate bonds, which are traded less frequency than many other assets such as stocks, similarity computation can be of great importance in reducing tracking error, transaction costs, as well as to replace bonds with specific themes such as the ones having higher ESG scores. 

Most of the research literature on this topic though has been focused on unsupervised clustering which has several drawbacks. The unsupervised methods may not provide information on whether we have sufficient input features to compute the similarity accurately, nor do they directly yield feature importance or provide objective metrics to evaluate the results. In the present work, we started by arguing that supervised similarity, which mostly is focused on distance metric learning (DML) from the given data, is a more appropriate way to compute similarity among assets. Then, we used Random Forest (RF) based DML which is in turn based on proximity computation. In a nutshell, after a RF is trained on the training dataset, one can compute similarity between two data-points by counting the number of times the pair falls in the same leaf node out of all the trees in the forest, called RF proximity. We used variables listed in \ref{tab:features_list} as input features and the target variables to train RF for corporate bonds and used two versions of proximities to compute similarities among corporate bonds.

To evaluate similarities computed using RF proximities and using traditional distance metrics (such as Euclidean and Gower), we proposed a novel KNN-based metric which takes in model specific distance metric and predicts the target variable based on KNN with respect to the corresponding distance metric. Compared with the traditional distances, the distance metrics based on RF proximities perform better.

In the future, we will investigate effects of trading corporate bonds using the similarity learning method investigated in the present work on performance of actively managed funds as well as on tracking error of index funds. 

There also exist other definitions of RF proximities (see Ref.~\cite{rhodes2022geometry} for a recent review on different proximity definitions) which we plan to investigate to identify the most appropriate proximity definition for corporate bonds. In addition, we also plan to use other tree-based methods such as extreme trees and Gradient-boosted machines to learn and extract similarity in a supervised fashion.

\section*{Acknowledgement}
The views expressed here are those of the authors alone and not of BlackRock, Inc.

\bibliographystyle{unsrt}
\bibliography{main}

\appendix
\section{ Permutation Feature importance}
\label{appendix:a}
Permutation feature importance is a model evaluation technique that can be used with any fitted estimator and when the input data is tabular. This implementation is readily available in scikit-learn package. It is defined as the decrease in the model score when a single feature value is randomly shuffled. By doing this we can break the relationship between the feature and the target and in turn explain the drop in the model score. This process can be repeated multiple times with different permutations of the feature. 

\begin{figure}[htbp]
\centering
\includegraphics[width=\linewidth]{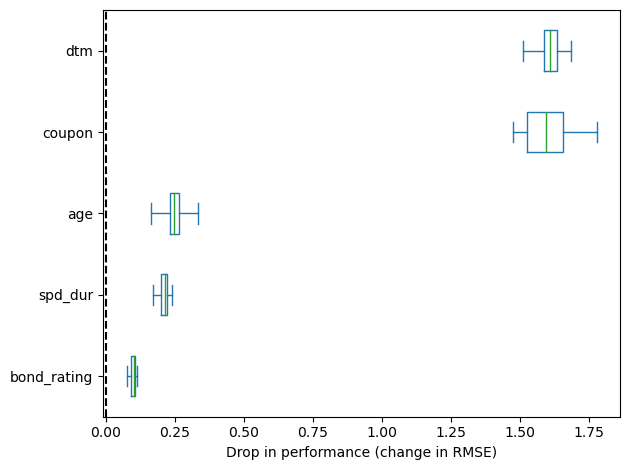}
\caption{Random Forest permutation feature importance}
\label{fig:feature_imp}
\end{figure}

Figure \ref{fig:feature_imp} shows the permutation feature importance for the top 5 features which contribute to the model performance. The x-axis shows the scale to which the RMSE would change (increase, in turn decreasing the model performance) and the y-axis provide the respective feature name.

\end{document}